\documentclass[letterpaper]{article}

\usepackage{natbib,alifeconf}

\usepackage[T1]{fontenc}

\usepackage{subcaption}

\usepackage{amsmath}

\usepackage[final]{hyperref}

\usepackage[]{algorithm2e}
\usepackage{tabularx}

\graphicspath{{annexes/}}

\renewcommand{\vec}[1]{\mathbf{#1}}

\title{APOGeT: Automated Phylogeny over Geological Time-scales}
\author{Kevin Godin-Dubois, Sylvain Cussat-Blanc \and Yves Duthen \\
\mbox{}\\
University of Toulouse, IRIT - CNRS UMR 5505, 2 rue du Doyen Gabriel Marty, 31042 Toulouse, France \\
\{kevin.dubois, sylvain.cussat-blanc, yves.duthen\}@irit.fr}
\makeatletter
\hypersetup{
 pdfinfo={
  Author={Kévin Godin-Dubois},
  Title={\@title}
 }
}
\makeatother

\begin{document}

\maketitle

\begin{abstract}
To tackle the challenge of producing tractable phylogenetic trees in contexts where complete information is available, we introduce APOGeT: an online, pluggable, clustering algorithm for a stream of genomes.
It is designed to run alongside a given experimental protocol with minimal interactions and integration effort.
From the genomic flow, it extracts and displays species' boundaries and dynamics.
Starting with a light introduction to the core idea of this classification we discuss the requirements on the genomes and the underlying processes of building species' identities and managing hybridism.
Though stemming from an ALife experimental setting, APOGeT ought not be limited to this field but could be used by (and benefit from) a broader audience.
\end{abstract}

Manual clustering of a biological population through various measurements (morphological, behavioral, molecular) allows one to define species.
Furthermore, by looking back into the past, through the fossil record, long-dead trends of genetic uniqueness can be identified by performing the same kind of clustering. These are then chained together by drawing blurry ancestry lines between successive ``snapshots''.
 
But Artificial Life is not constrained by such limitations.
Indeed in computer-embedded evolutionary processes one has access to much more information: the whole genealogy itself.
This, however, is impractical for most purposes and extracting, a posteriori, the relevant aggregate information from such an amount of data is a daunting task.
 
Instead we propose to filter out individual species directly from the stream of created/destroyed genomes thus alleviating the burden of post-processing.
In this work we use the \emph{biological} species concept, as described in \citep{Singh2012}, which is a ``group of potentially interbreeding natural population reproductively isolated from other such groups''.
This pose, however, a problem of its own: consider a single individual, which cannot be very different from either its parents or direct descendants.
There ought not be a point in this stream where one could rightfully say: ``here is a new species''.
We address this problem by relying on the concept of representatives, i.e. a small collection of genomes (\emph{R-set}) that best describes the diversity of their species in a given genetic space $\vec G$.
 
\begin{figure}[t]
 \def\s{.78}
 \begin{subfigure}[b]{.22\textwidth}
  \centering
  \includegraphics[scale=\s]{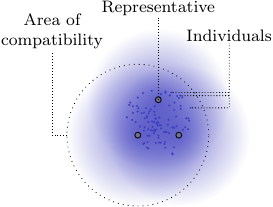}
  \caption{Low variability}
  \label{fig:envlp:uniform}
 \end{subfigure}
 \begin{subfigure}[b]{.22\textwidth}
  \centering
  \includegraphics[scale=\s]{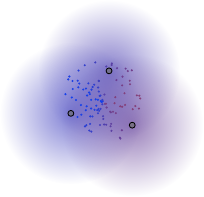}
  \caption{Diversity increasing}
  \label{fig:envlp:diversity}
 \end{subfigure}
 
 \begin{subfigure}[b]{.22\textwidth}
  \centering
  \includegraphics[scale=\s]{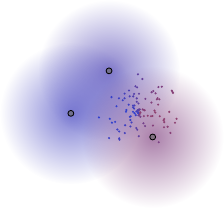}
  \caption{Low intra-compatibility}
  \label{fig:envlp:stretched}
 \end{subfigure}
 \begin{subfigure}[b]{.22\textwidth}
  \centering
  \includegraphics[scale=\s]{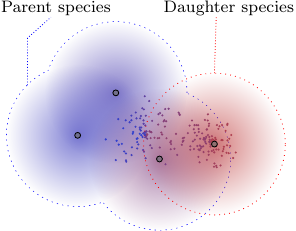}
  \caption{Cladogenesis}
  \label{fig:envlp:genesis}
 \end{subfigure}
 \caption{Implicit species boundaries by means of an R-set}
\end{figure}

Assuming that we start from a relatively homegeneous species as depicted in figure \ref{fig:envlp:uniform}, we can define, for each representative, the region of genetic space with which it is compatible (according to some criteria).
The species volume can then be derived, as the region where there is sufficient overlap between the members of the R-set.
As diversity starts to increase this collection of representatives will be updated to reflect the changes in ``typical'' features (fig. \ref{fig:envlp:diversity}). 

There comes a time, however, where the overlapping regions of compatibility will start to wear thin (fig. \ref{fig:envlp:stretched}), preventing further increase in the diversity of the species. From this point on, diverging genomes will no longer be a good match and will be assigned to a daughter species (fig. \ref{fig:envlp:genesis}).

\subsection{Prerequisites}

For this clustering we require only a handful of characteristics from the underlying genomes, namely:

\begin{tabular}{>{\bf}rl}
        events & Birth/Death (or equivalent) \\[-.1cm]
    & \small \emph{insertion into/suppression from the tree} \\
     genealogy & Access to their parents identity \\[-.1cm]
    & \small \emph{direct access to the candidate species} \\
 compatibility & $C: \vec{G}^2 \to [0,1]$ \\[-.1cm]
    & \small \emph{match between two genomes in $\vec G$}
\end{tabular}

\subsection{Species affectation}

Given a genome $g$, we can easily test whether it belongs in its parents' species $S$, represented by $R=\{r_1,\dots,r_K\}$, by computing its compatibility with each $r_i$.
The averaged figure is matched against a user specified threshold $T$ to determine whether or not $g$ will be inserted in $S$.
When the test fails, the procedure is repeated for every daughter species until a match is found.
If no such case occurs, a new subspecies of $S$ is created with $g$ as its sole representative.

Additionally, each insertion of a genome $g$ into a species with a full representative set $R$ must test whether there exists a $r_i$ that is a worse representative than $g$. The objective is to maintain $R$ as the set of individuals that are the most different from one another. The current implementation tests whether $g$ has lower compatibility with at least one member of the set (with the exclusion of $r_i$) while not being ``closer'' to the rest of the representatives.

The detail-oriented reader might have inferred that a contradiction stems from this representation: it is not only possible, but actually guaranteed, that some individuals will end up in different species from their parents due to the arbitrary cut-off.
While this is a necessary by-product of packaging continuous dynamics into sharply defined compartments, it ought not be a problem in itself, as long as this classification does not impact the individuals' dynamics.

\subsection{Hybridism}

Handling of hybridism is, thus, a serious concern in this model.
Indeed, from the previous point, we can expect to see mating between close genomes, unaware of any arbitrary species ``barrier'' between them.
Experimenters might also be interested in investigating the impact of various types of hybridism, especially in the case of plants.

We address this problem by now considering that, for a $g$, we might have two candidates species for insertion.
The procedure needs only be slightly altered by first comparing which of these have the highest average compatibility with $g$.
We then assume that the genome will belong in the phylogenetic subtree of this most similar species and restart the procedure from this point.
Moreover, we keep track of each individual contribution to the ``gene pool'' of a given species $S$ so that we can define a \emph{major contributor}, i.e. the species $P \neq S$ that provided the most genetic material.
This gives rise to a robust definition of ``parent species'' in a context where genes flow unregulated.

\section{Conclusion}

\begin{figure}[t]
 \def\w{.2\textwidth}
 \def\s{.025}
 \hfill
 \begin{subfigure}{\w}
  \centering
  \includegraphics[scale=\s]{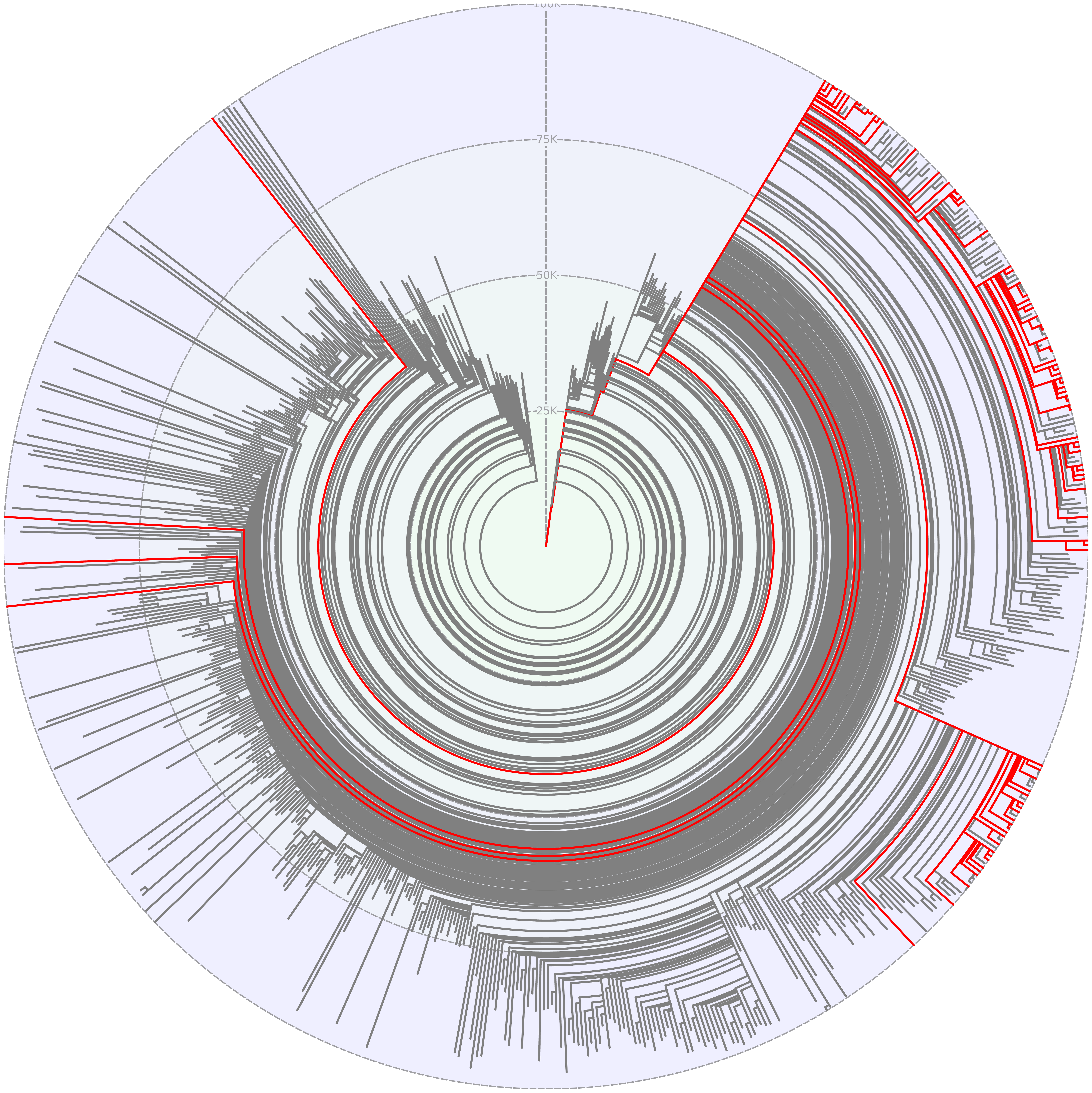}
  \caption{Full tree}
  \label{fig:all}
 \end{subfigure}
 \hfill
 \begin{subfigure}{\w}
  \centering
  \includegraphics[scale=\s]{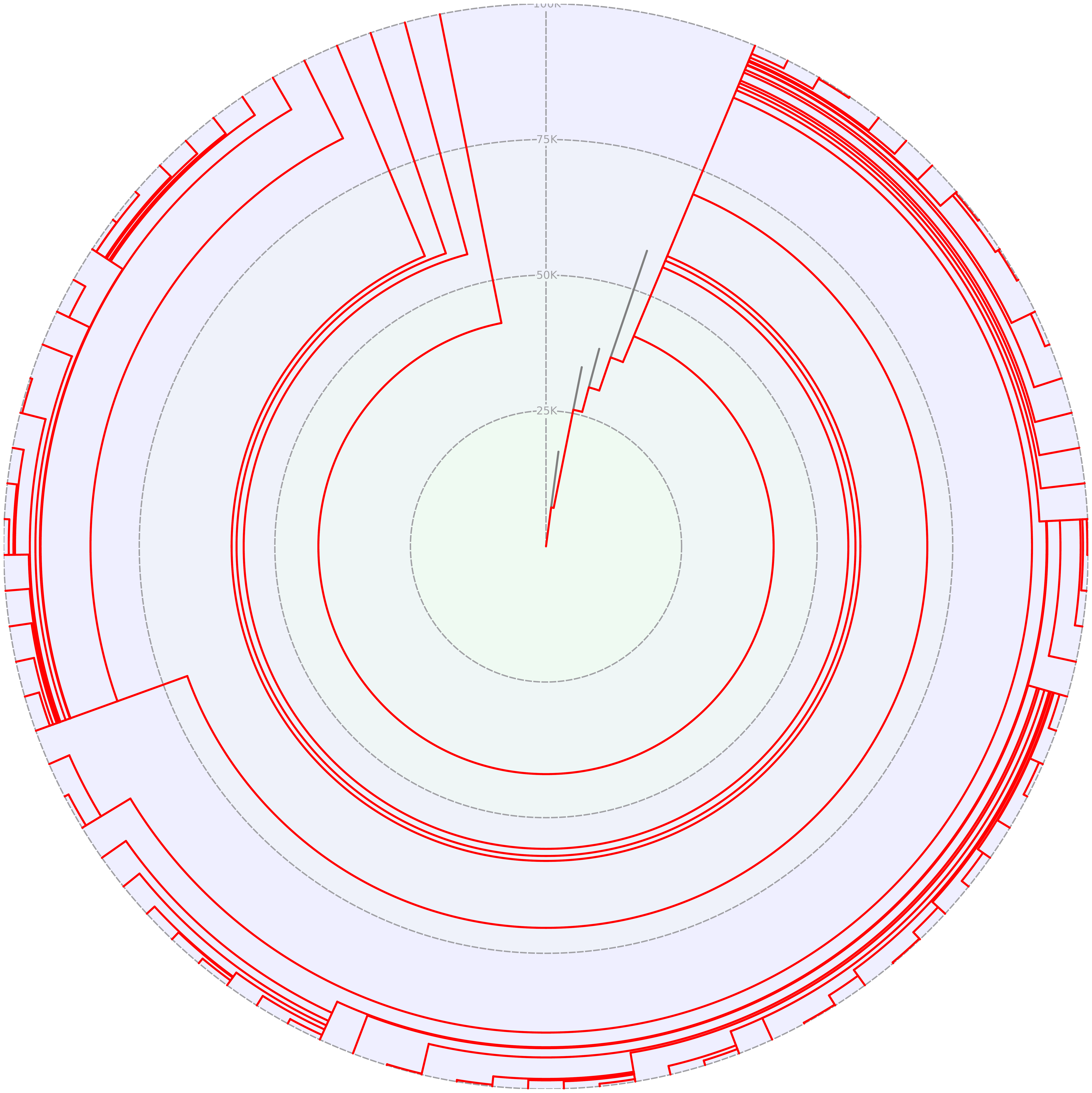}
  \caption{Survivors only}
  \label{fig:survivors}
 \end{subfigure}
 \hfill
 
 \hfill
 \begin{subfigure}{\w}
  \centering
  \includegraphics[scale=\s]{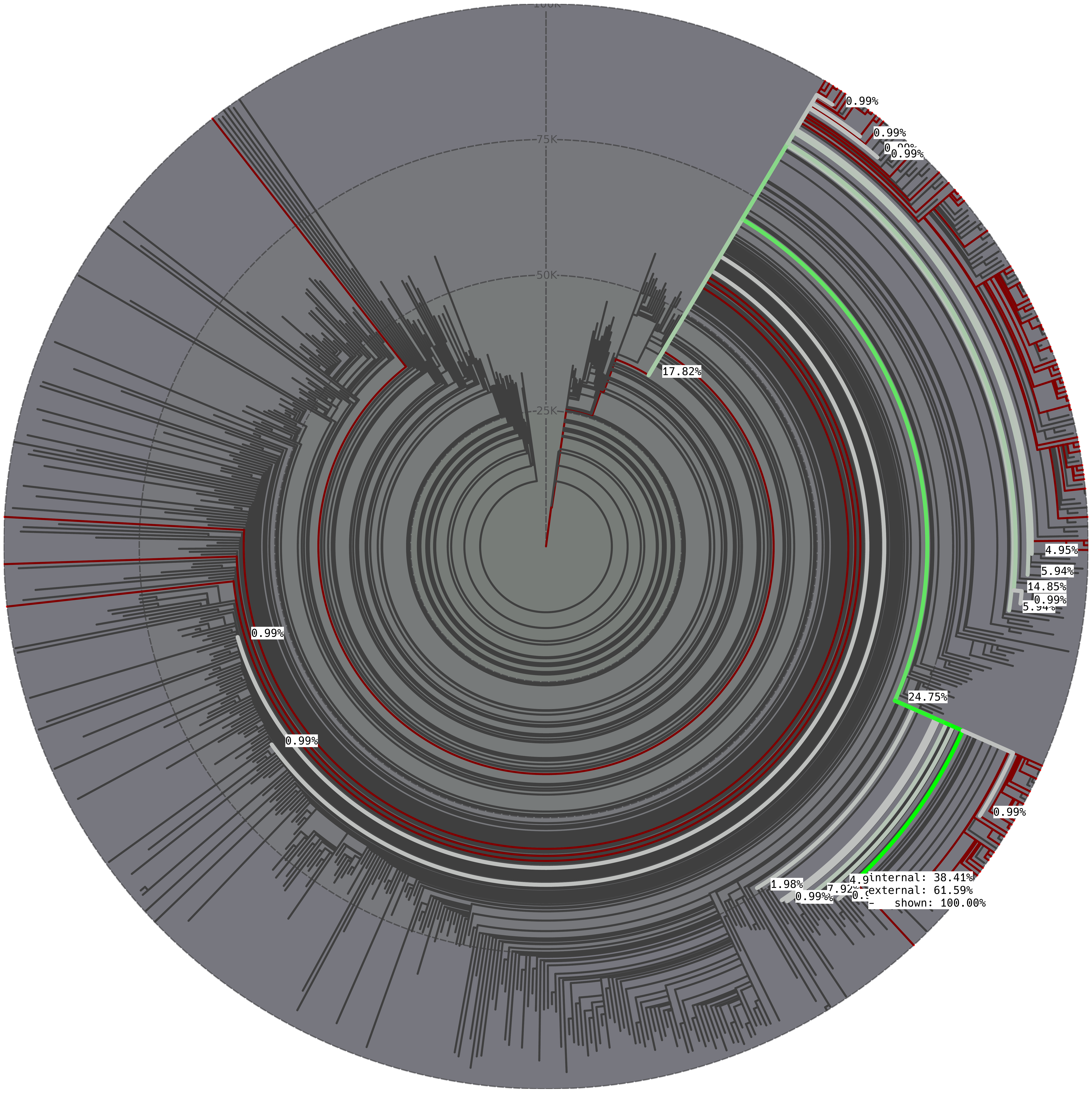}
  \caption{Hybridism}
  \label{fig:contributors}
 \end{subfigure}
 \hfill
 \begin{subfigure}{\w}
  \centering
  \includegraphics[scale=\s]{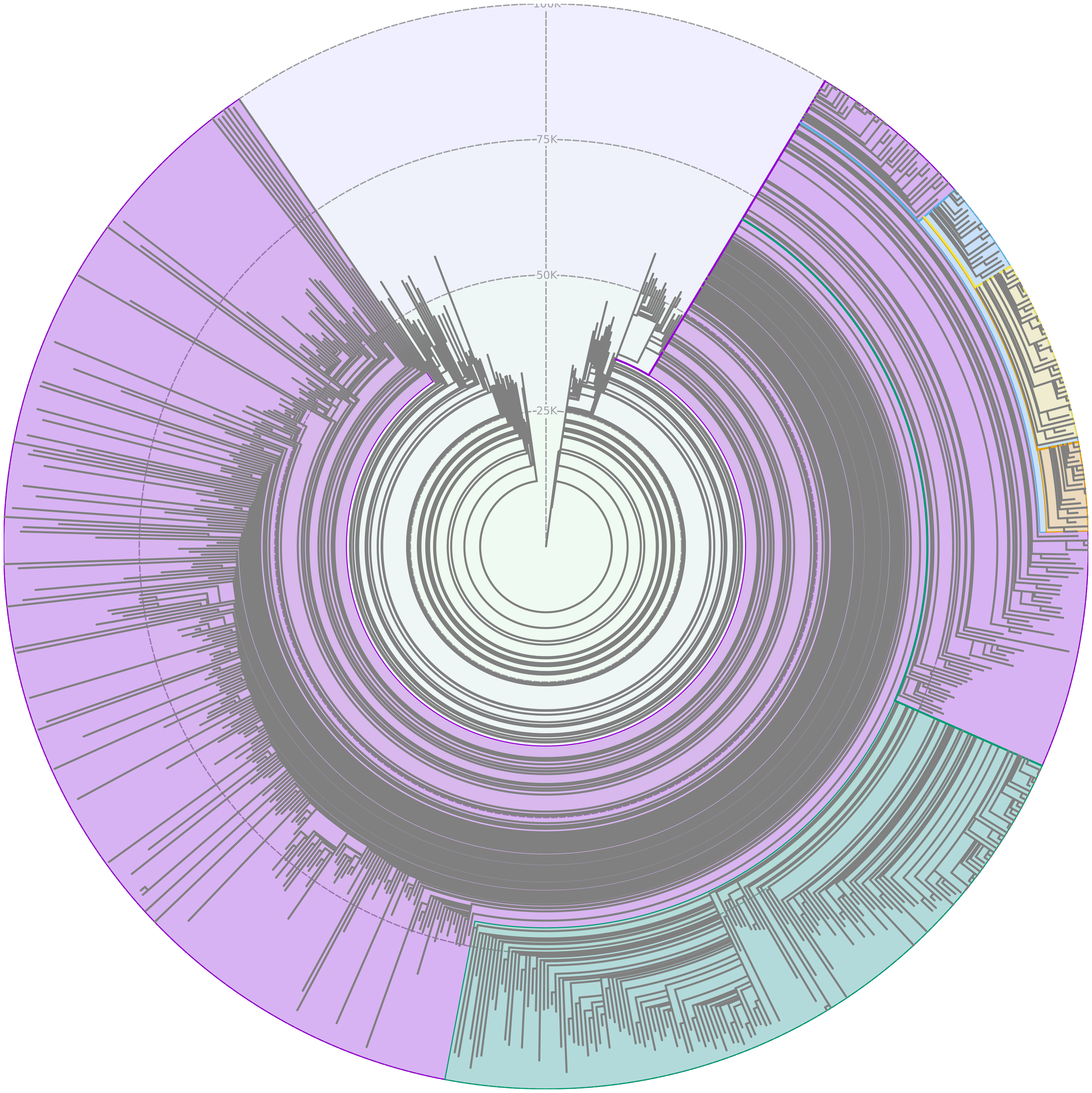}
  \caption{Taxonomic levels}
  \label{fig:tracking}
 \end{subfigure}
 \hfill
 \caption{Examples of visualisations}
\end{figure}

In addition to the core, phylogeny-computing component, this tool comes with a user interface to facilitate extraction of gross information (e.g. figures \ref{fig:all}, \ref{fig:survivors}) during the simulation. Further visualisations include, but are not limited to, the graph of hybridism for a given species (fig. \ref{fig:contributors}) and defining species groups (genus, family, ... see fig. \ref{fig:tracking}).

Once the C++ repository at \url{https://github.com/kgd-al/APOGeT} is retrieved, integrating this tool with a given experiment requires setting two parameters ($T$, $K$) and devising $C$.
The ``belonging'' threshold $T$ is of utmost importance as it will have crucial impact on the clustering fidelity.
The value, however, is not domain dependent and initial experiments show that $T=.25$ provides exploitable data.
The representatives set size $K$, on the other hand, is loosely tied to the dimensionality of $\vec G$: if set too low, the diversity of the underlying genomic space will be poorly encapsulated, while too high a value will result in overfitting and a heavier CPU/memory footprint.

Finally, designing a good compatibility metric $C$ might prove difficult, depending on the task and genome model, but our main concern is the sensitivity of our tool to ``noise'' in the input stream.
That is, a number of species produced by the clustering procedure are of no relevance to the long term dynamics of the simulation but must still be tested for belonging in each genomic insertion, in case of a late ``kick-off''.

As mentioned earlier, we devised this tool while working on ALife experiments, see \citep{GodinDubois2019b}, and thus it is biased towards specific evolutionary pattern.
However, given the limited set of constraints we need to impose on the genomic stream, it should be relatively straightforward to extent to other areas of the computer sciences including traditional optimization through evolutionary algorithms.
Manipulation of real biological data, however, would prove more difficult due to limited availability and the complexity of devising a robust compatibility metric.

\footnotesize
\bibliographystyle{apalike}
\bibliography{main}

\begin{thebibliography}{}

\bibitem[{Godin-Dubois} et~al., 2019]{GodinDubois2019b}
{Godin-Dubois}, K., {Cussat-Blanc}, S., and Duthen, Y. (2019).
\newblock Speciation under {{Changing Environments}}.
\newblock In {\em {{ALIFE}} 19}, volume~31, pages 349--356, Cambridge, MA. MIT
  Press.

\bibitem[Singh, 2012]{Singh2012}
Singh, B.~N. (2012).
\newblock Concepts of species and modes of speciation.
\newblock {\em Current Science}, 103(7).

\end{thebibliography}

\end{document}